\documentclass[a4paper,11pt]{article}
\usepackage{jheppub}
\usepackage{upgreek}
\usepackage{subcaption}
\usepackage{tikz-feynman}
\usepackage{amsmath}
\usepackage{slashed}
\usepackage{cleveref}
\usepackage{multirow}
\usepackage{orcidlink}
\usepackage{tcolorbox}
\usepackage[table]{xcolor}
\usepackage{adjustbox}

\newcommand{\lnp}{\Lambda^{}_{\text{NP}}}

\newcommand{\mh}{m_{h}}

\newcommand{\mphi}{m_{\Phi}^{}}
\newcommand{\Gphi}{\Gamma_{\phi}}

\newcommand{\Trh}{T_{\text{rh}}}
\newcommand{\Tmax}{T_{\text{max}}}
\newcommand{\TeV}{\text{TeV}}
\newcommand{\TFO}{T_\text{FO}}
\newcommand{\SM}{\text{SM}}
\newcommand{\GeV}{\text{GeV}}
\newcommand{\MeV}{\text{MeV}}
\newcommand{\mpl}{m_{\rm pl}}

\newcommand{\HI}{\mathcal{H}_\text{I}^{}}
\newcommand{\sphin}{\sigma^{\text{SI}}_{\Phi\text{N}}}
\newcommand{\lphi}{\lambda_{\Phi}^{}}
\newcommand{\lphiH}{\lambda_{\Phi\text{H}}^{}}
\newcommand{\lphiHp}{\lambda_{\Phi\text{H}}^{\prime}}
\newcommand{\lphiHpp}{\lambda_{5}}

\newcommand{\mut}{\mu_{3}}
\newcommand{\eq}{eq\,.~}

\newcommand{\eqs}{eqs\,.~}

\newcommand{\fig}{fig\,.~}

\definecolor{deepmagenta}{rgb}{0.8, 0.0, 0.8}
\definecolor{mediumtealblue}{rgb}{0.0, 0.33, 0.71}
\definecolor{warmblack}{rgb}{0.0, 0.26, 0.26}
\definecolor{bostonuniversityred}{rgb}{0.8, 0.0, 0.0}
\definecolor{junglegreen}{rgb}{0.16, 0.67, 0.53}
\definecolor{lightcornflowerblue}{rgb}{0.6, 0.81, 0.93}
\definecolor{mypink1}{rgb}{0.858, 0.188, 0.478}
\definecolor{mypink2}{RGB}{219, 48, 122}
\definecolor{mypink3}{cmyk}{0, 0.7808, 0.4429, 0.1412}
\definecolor{mygray}{gray}{0.2}
\definecolor{ForestGreen}{RGB}{34,139,34}

\usepackage[font=small]{caption}
\usepackage[font={small},labelfont={bf}]{caption}
\usepackage{url}
\definecolor{MyDarkBlue}{rgb}{0.1, 0.1, 0.8}
\definecolor{SBlue}{rgb}{0.2, 0.4, 0.7} 
\definecolor{MyLightBlue}{rgb}{0.22,0.51,0.9}
\definecolor{MyGreen}{rgb}{0.0, 0.5, 0.0}
\definecolor{BrickRed}{rgb}{0.8, 0.25, 0.33}
\usepackage{hyperref}
\hypersetup{colorlinks, citecolor=BrickRed,linkcolor=MyDarkBlue, urlcolor=MyGreen}
\title{Complex scalar dark matter with effective Higgs portals beyond radiation domination}
\author[a,b]{Manimala Mitra\orcidlink{0000-0002-8032-5125}, }
\emailAdd{manimala@iopb.res.in}
\author[a,b]{Dipankar Pradhan\orcidlink{0000-0002-2450-6677}, }
\emailAdd{dipankar.pradhan@iopb.res.in}
\author[a,b]{Subham Saha\orcidlink{0009-0009-1183-3271}}
\emailAdd{subham.saha@iopb.res.in}
\affiliation[a]{Institute of Physics, Sachivalaya Marg, Bhubaneswar, Odisha 751005, India}
\affiliation[b]{Homi Bhabha National Institute, BARC Training School Complex, Anushakti Nagar, Mumbai 400094, India}

\abstract{
The increasingly stringent bounds on the Higgs-portal coupling, arising from dark matter (DM) direct-detection searches, confront the minimal renormalizable complex scalar DM scenario with thermal production, where freeze-out occurs in the standard radiation-dominated era. This limitation can be alleviated by introducing a dimension-5 Higgs-portal operator in the minimal renormalizable complex scalar DM model and/or by modifying the standard cosmological history of the Universe. In this article, we analyze complex scalar DM production in both the reheating and radiation-dominated epochs within an effective field theory (EFT) framework. While both scenarios exhibit sizeable regions of parameter space consistent with existing constraints, freeze-out during reheating opens up additional viable regions that are otherwise ruled out by DM overabundance in the radiation-dominated scenario. Notably, the renormalizable Higgs-portal coupling is constrained by relic density, direct- and indirect-detection limits, whereas the EFT coupling associated with the dimension-5 operator is constrained by relic density and indirect-detection bounds arising from DM semi-annihilation. We further study the production cross section of complex scalar DM at hadron and lepton colliders.
}
\preprint{IOP/BBSR/2025-03}

\begin{document}
\maketitle
\flushbottom
\section{Introduction}
Despite strong theoretical and observational \cite{Cirelli:2024ssz, Bertone:2004pz} motivations for dark matter (DM), its identity and fundamental properties remain elusive. Among the myriad proposed candidates, spin-0 scalar DM \cite{McDonald:1993ex} provides the simplest framework for exploring its dynamics and interactions. The production of scalar DM during the Standard Model (SM) radiation-dominated era, particularly when the reheating temperature is much larger than the freeze-out temperature, has extensively been studied in literature \cite{Boehm:2020wbt, Boehm:2003hm, Han:2015hda, Duerr:2015aka, Arcadi:2019lka}. Scalar DM is often overproduced once constraints from direct and indirect detection experiments and collider searches are taken into account.
This can be evaded if the minimal renormalizable setup is being extended by additional dark-sector states \cite{Deshpande:1977rw,LopezHonorez:2006gr,Bhattacharya:2024nla,Lahiri:2024rxc,Bhattacharya:2025mlg,Bhattacharya:2024jtw,Habibolahi:2022rcd, Cao:2007fy, Hambye:2009pw, Bhattacharya:2013hva} or by Effective Field Theory (EFT) operators \cite{Bauer:2016pug, Fitzpatrick:2012ix, Cao:2009uw, Alanne:2017oqj} involving DM and SM fields. Alternatively, a modification of the standard cosmological history, such as entropy dilution \cite{Belanger:2024yoj, Gelmini:2006pq, Arbey:2021gdg, Drees:2006vh} during the reheating era also serve as a viable framework to overcome this constraint.

Among scalar DM candidates, the SM gauge-singlet real scalar offers the simplest realization, whereas complex scalar dark matter (CSDM) is a viable alternative.
A CSDM, stabilized by a $\mathcal{Z}_3$ symmetry \cite{Belanger:2012zr, Gonderinger:2012rd, Barger:2008jx}, and interacting with the Higgs through renormalizable portal coupling is severely constrained by direct detection (DD) experiments.
Based on the recent $\rm LZ\text{-}2025$ \cite{LZ:2024zvo} results, the spin-independent (SI) DM-nucleon scattering cross-section larger than $[10^{-46} -10^{-48}]\,\rm cm^2$ are excluded for DM masses up to $10~\TeV$. 
This imposes severe constraint on the CSDM parameter space, except around the Higgs resonance.
Even if the next-to-leading-order (NLO) corrections of the Higgs portal coupling are being included \cite{EscuderoAbenza:2025cfj}, the bound almost remain the same.
Additional constraints in this region arises from Fermi-LAT observations of DM self-annihilation \cite{Armand:2022sjf, Fermi-LAT:2016uux, Fermi-LAT:2016afa}. 
The severe constraints in this minimal setup arise because the same portal coupling controls both the DM relic abundance and direct detection cross section. Once constraints from direct detection are imposed, the DM relic density is overabundant in most of the parameter space.

A promising approach to enlarge the viable parameter space of the CSDM model is by introducing additional dimension (dim)-5 interaction term involving CSDM and SM Higgs into the theory, that does not contribute to CSDM–nucleon scattering but plays a crucial role in the DM freeze-out process, and hence DM relic density can be satisfied. In this work, we study this $d=5$ EFT extension focussing on scenarios, when DM freeze-out occurs in the radiation dominated era, and during reheating. We show that compared to the standard scenario of a instantaneous reheating, a much larger parameter space is allowed by the DM relic density constraint if the freeze-out occurs during reheating.
We further constrain the model parameter space using gamma-ray observations from Fermi-LAT, focusing on limits derived from DM self- and semi-annihilation processes. Since the semi-annihilation rate depends on the Wilson coefficient of the DM-EFT operator, we ensure the validity of the EFT description, which imposes constraints on the DM mass, the new-physics scale, and the maximum temperature of the thermal bath. After imposing relic-density requirements as well as direct and indirect detection bounds, the DM production cross sections at future colliders \cite{FCC:2018vvp, FCC:2025uan, FCC:2025lpp, ILC:2007oiw, ILC:2013jhg, ILC:2007vrf, ILC:2007bjz,ZurbanoFernandez:2020cco} are found to be very small $\sim 10^{-3}\,\mathrm{fb}$, rendering discovery through conventional cut-based analyses highly non-trivial.

The paper is organised as follows. In \Cref{sec:model}, we present the CSDM framework, in which the dark sector interacts with the SM through the Higgs portal. In \Cref{sec:DM}, we discuss the experimental constraints on the parameter space consistent with the observed relic density and explore prospects for detection in future DM experiments and evaluated DM production cross section in future colliders. Finally, the summary and conclusions are given in \Cref{sec:summary}.

\section{The Model}
\label{sec:model}
In this article, we extend the renormalizable Higgs-portal CSDM model by introducing a dim-5 CSDM–Higgs EFT operator, $\mathcal{O}_5: \Lambda_{\rm NP}^{-1}\Phi^3(H^\dagger H)$. In contrast, in the case of real scalar and vector boson DM, the leading gauge-invariant DM–EFT operator arises only at dim-6, which is always suppressed. Moreover, for fermionic DM, the leading operator arises at dim-5 and is tightly constrained by DD and ID constraints \cite{Arcadi:2024wwg}.
We explore both the possibilities, standard thermal production of CSDM in the radiation-dominated era, as well as, during reheating. The latter scenario is augmented with two new parameters: the inflaton decay width and the expansion rate at the end of inflation. Upto dim-5, the Lagrangian corresponds to a Higgs-portal SM gauge-singlet CSDM field $(\Phi)$, which transforms under a $\mathcal{Z}_3$ symmetry as $\Phi \to e^{i 2\pi/3} \Phi$ is \cite{Wu:2016mbe, Gonderinger:2012rd, Belanger:2012zr, Hektor:2019ote}:
\begin{gather}
\begin{split}
\mathcal{L}_{}\,\supset\,|\partial_{\mu}\Phi |^2-\mu_{\Phi}^2|\Phi|^2-\lphi|\Phi^*\Phi|^2-\frac{1}{2}\mu_{3}\left[\Phi^3+\left(\Phi^{*}\right)^3\right] -\lphiH |\Phi|^2H^{\dagger}H \\ 
-\Lambda_{\rm NP}^{-1}\left[\Phi^3+\left(\Phi^{*}\right)^3\right]\left(\lphiHp H^{\dagger}H + \lphiHpp|\Phi|^2\right)+\mathcal{O}(\Lambda_{\rm NP}^{-2})\,,\hspace{2cm}
\label{eq:model}
\end{split}
\end{gather}
We follow the notation of \cite{Hektor:2019ote} in defining the Lagrangian.
Other higher dimensional operators would always give a suppressed contribution compared to the dim-5 operator. The dim-5 self-interaction term is only relevant for low masses, where $\mphi \sim \mathcal{O}~(\text{MeV})$, and it contributes dominantly to the DM relic abundance through the $3\Phi \to 2\Phi$ annihilation process \cite{Cervantes:2024ipg}. Therefore, we remain agnostic about $\lphiHpp$ throughout the rest of this work.
The Higgs field acquires a non-zero vev after spontaneous Electroweak Symmetry Breaking (EWSB), $\langle H\rangle \to v/\sqrt{2}$ with $v=246~\GeV$, and the CSDM mass is modified as $m_\Phi^2=\mu_{\Phi}^2+\dfrac{1}{2}\lphiH v^2$. 

The search for Higgs invisible decays by the ATLAS detector, using $139~\mathrm{fb}^{-1}$ of proton-proton collision data at a centre-of-mass energy of $\sqrt{s} = 13~\TeV$ at the LHC, combined with results obtained at $\sqrt{s} = 7~\TeV$ and $\sqrt{s} = 8~\TeV$, sets an observed upper limit of $0.107$ on the Higgs boson branching ratio to invisible final states at the $95\%$ confidence level \cite{ATLAS:2023tkt}. A combined analysis of $\sqrt{s} = 7$, $8$, and $13~\TeV$ proton–proton collision data at the CMS detector sets an observed $95\%$ confidence level upper limit of $0.15$ on $\mathcal{B}(h \to \mathrm{inv})$ \cite{CMS:2023sdw}:
\begin{equation}
\mathcal{B}(h\to \text{invisible})<
\begin{cases}
0.107\quad\text{ATLAS [95$\%$ CL]}\,.\\
0.15\qquad\text{CMS [95$\%$ CL]}\,.
\end{cases}
\label{eq:higgs-invisible}
\end{equation}
In this model framework, the Higgs can decay into CSDMs, $h\to \Phi\Phi^*\,(\Phi^3/\Phi^{*3})$, if $\mh\gtrsim 2\mphi\,(3\mphi)$. However, the constraint on the Higgs-portal coupling from Higgs invisible decays is always weaker \cite{ATLAS:2023tkt} than the constraint from the spin-independent CSDM-nucleon scattering cross section measured by $\rm LZ-2025$. Therefore, we do not impose the Higgs invisible decay constraint, rather consider only the direct detection bound discussed in \cref{sec:relic-DD}.

\section{Dark Matter Phenomenology}
\label{sec:DM}

In the reheating era, the inflaton field $(\phi)$ decays into SM radiation with decay width $\Gphi$, and its energy density evolution is defined by the Boltzmann equations (BEQs) \cite{Gelmini:2006pw, Belanger:2024yoj}:
\begin{gather}
\label{beq:inflaton}
\dfrac{d\rho_{\phi}}{dt}+3\mathcal{H}\rho_{\phi}=-\Gphi\rho_{\phi}{\color{blue}\footnotemark[1]}\,,\\
\dfrac{d\rho_{r}}{dt}+4\mathcal{H}\rho_{r}=+\Gphi\rho_{\phi}\,,
\label{beq:radiation}
\end{gather}
\footnotetext[1]{For a monomial potential $(V(\phi)=\lambda \phi^n)$, the Boltzmann equation becomes $\dfrac{d\rho_{\phi}}{dt} + 3\mathcal{H}(1+\omega_{\phi})\rho_{\phi} = -\Gamma_\phi \rho_{\phi}$,
where $\omega_{\phi} = p_{\phi}/\rho_{\phi} = (n-2)/(n+2)$ is the effective equation-of-state parameter of the inflaton.}
where $\rho_{\phi}~(\rho_r)$ is the inflaton (radiation) energy density, and, equivalently for entropy, \eq\eqref{beq:radiation} reduces to,
\begin{gather}
\dfrac{d s}{dt}+3\mathcal{H}s=+\dfrac{\Gphi\rho_{\phi}}{T}\,,
\label{beq:entropy}
\end{gather}

The entropy density ($s$) is defined as a function of the SM radiation temperature, T,
\begin{align}
s=\dfrac{2\pi^2}{45}h_{\rm eff}(T)T^3\,,
\end{align}
where $h_{\rm eff}$ is the number of relativistic degrees of freedom contributing to the entropy density \cite{Drees:2015exa}, and the Hubble expansion rate ($\mathcal{H}$) is given by
\begin{align}
\mathcal{H}^2=\dfrac{8\pi}{3}\dfrac{\rho_{\phi}+\rho_R}{m_{\rm pl}^2}\,,
\end{align}
where $\mpl\simeq 1.2 \times 10^{19}\rm ~GeV$ is the planck masss and $\rho_R$ denotes the Standard Model energy density and is given by,
\begin{align}
\rho_R=\dfrac{\pi^2}{30}g_{\rm eff}(T)T^4\,,
\end{align}
where $g_{\rm eff}$ is the number of relativistic degrees of freedom contributing to $\rho_R$.

The reheating temperature ($\Trh$) is defined as the temperature where $\rho_{\phi}(\Trh)=\rho_{R}(\Trh)$ and estimated implicitly by the equality $\mathcal{H}(\Trh)=\Gphi$ which implies \cite{Belanger:2024yoj},
\begin{align}
T_{\rm rh}^2=\dfrac{3}{2\pi}\sqrt{\dfrac{5}{\pi g_{\rm eff}(\Trh)}}\mpl\Gphi\,,
\end{align}
and the maximum temperature $\Tmax$ reached during reheating by the thermal bath, which can be estimated by \cite{Barman:2021ugy, Bernal:2019mhf}
\begin{align}
T_{\rm max}^4=\dfrac{15}{2\pi^3 g_{\rm eff}(\Tmax)}\left(\dfrac{3}{8}\right)^{8/5}\mpl^2\Gphi\HI\,.
\label{eq:tmax}
\end{align}
We have rewritten \eqs\eqref{beq:inflaton} and \eqref{beq:entropy} in terms of the cosmic scale factor $a$. To solve the BEQs, we use the initial conditions at the onset of reheating, $t = t_I$, corresponding to the scale factor $a_I \equiv a(t_I) = 1$. At this time, the radiation energy density is $\rho_R(t_I) = 0$, and the inflaton energy density is $\rho_{\phi}(t_I) = \frac{3}{8\pi} \mpl^2 \mathcal{H}_{\rm I}^2 \neq 0,$ where $\HI$ denotes the inflationary scale.
The latter has an upper bound of $2.5 \times 10^{-5} \mpl$ \cite{Planck:2018jri, BICEP:2021xfz}, derived from the non-observation of B-modes.
However, there also exist lower bounds on this inflationary parameter, derived from CMB observations and BBN. From the Planck measurements, the amplitude of scalar perturbations is found to be $A_s \approx 2.1\times 10^{-9}$, while the tensor-to-scalar ratio $r$ can, in principle, be arbitrarily small. The inflationary parameter is related to these quantities as $\HI \simeq \pi \mpl \sqrt{\frac{r\,A_s}{2}}$ \cite{Baumann:2009ds}. Therefore, inflation could occur at a much lower energy scale without contradicting the CMB data. However, for BBN to proceed normally, inflation must end early enough to allow reheating to produce the SM bath with temperature $\,\Trh \gtrsim 4\,\MeV$ \cite{Sarkar:1995dd, Hannestad:2004px,deSalas:2015glj}.

The evaluation of CSDM number density ($n_{\Phi}$) is governed by the Boltzmann equation,
\begin{align}
\dfrac{dn_{\Phi}}{dt}+3\mathcal{H}n_{\Phi}=-\langle\sigma v\rangle_{\Phi\Phi^*\to\rm SM~SM}\left(n_{\Phi}^2-n^{\rm eq^2}_{\Phi}\right)-\dfrac{1}{2}\langle\sigma v\rangle_{\Phi\Phi\to\rm \Phi^*~h}\left(n_{\Phi}^2-n^{}_{\Phi}n^{\rm eq}_{\Phi}{}\right)\,,
\label{eq:beq}
\end{align}
where $\rm SM=\{\text{leptons, quarks, Higgs},~W^{\pm}\text{ and Z}\}$, $\langle\sigma v\rangle_{\Phi\Phi^*\to\rm SM~SM}$ and $\langle\sigma v\rangle_{\Phi\Phi\to\rm \Phi^*~h}$ are the self and semi-annihilation cross-sections\footnote{The other annihilation channels, such as $3\Phi \to \Phi\, h\,,~2\Phi\,h\to\Phi\, h$ and $3\Phi \to h\, h$, are not included in \eq\eqref{eq:beq}, as these processes remain subdominant throughout the CSDM mass range considered.} of the CSDM, and $n_{\Phi}^{\rm eq}$ is the equilibrium number density of the CSDM. Finally, the DM relic abundance is followed by,
\begin{align}
\Omega_{\rm DM}h^2=2.744\times 10^8\dfrac{n_{\Phi}}{s}\bigg|_{T=T_0}\,,
\label{beq:dm}
\end{align}
where $T_0 \sim 2.35\times 10^{-13}~\mathrm{GeV}$ \cite{ParticleDataGroup:2024cfk} denotes present-day temperature of the universe.
The observed DM relic abundance reported by Planck \cite{Planck:2018vyg} is 
\begin{align}
\Omega_{\mathrm{DM}}^{\mathrm{obs}}h^2 = 0.1200 \pm 0.0012\,,
\end{align}
where $h$ is the reduced Hubble parameter $\rm H/100~km~s^{-1}~Mpc^{-1}$ with $\rm H=67.4\pm 0.5~km~s^{-1}~Mpc^{-1}$ \cite{Planck:2018vyg} being the current Hubble constant.
As it is well known that late-time entropy production from heavy particle decays enhances the post-freeze-out plasma entropy, diluting the comoving DM relic density and reducing its abundance below the standard thermal freeze-out value, thereby providing a minimal way to reconcile overabundant thermal DM relics with the observed DM relic density. In this article, the solution of BEQs [\eqs\eqref{beq:inflaton}, \eqref{beq:entropy} and \eqref{beq:dm}] and the estimations of $\Trh$ and $\Tmax$ are performed using the micrOMEGAs package \cite{Alguero:2023zol,Belanger:2024yoj}.
Importantly, it should be noted that $\Tmax$ must always satisfy the relation $\Tmax < \lnp$\footnote{The thermally average DM semi-annihilation cross section can be expressed as $\langle\sigma v\rangle \sim \mathcal{O}~(T/\lnp)$. Requiring the convergence of $\langle\sigma v\rangle$ implies $\lnp > \Tmax$, which also forbids the on-shell production of the heavy mediators. Nevertheless, the condition $\Tmax<\lnp$ is sufficient but not necessary \cite{Giudice:2000ex} for thermal DM freeze-out, since the relic density is controlled by the dynamics at the DM freeze-out temperature $\TFO~(\ll \Tmax)$.}; otherwise, it would violate the EFT validity criterion \cite{Garcia:2020eof}.

\subsection{Relic density and Direct detection}
\label{sec:relic-DD}
\begin{figure}[htb!]
\centering
\includegraphics[width=0.45\linewidth]{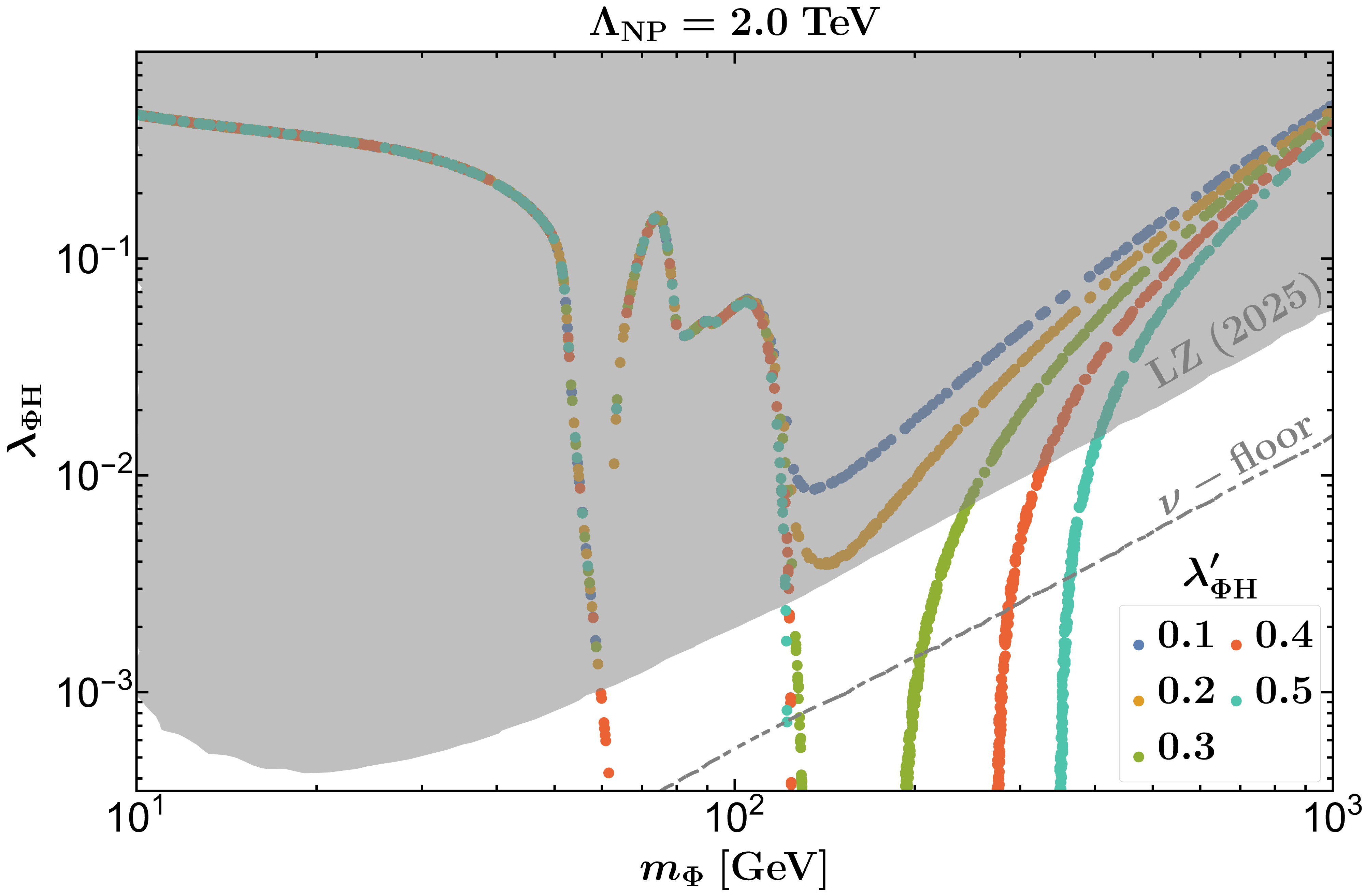}\quad
\includegraphics[width=0.5\linewidth]{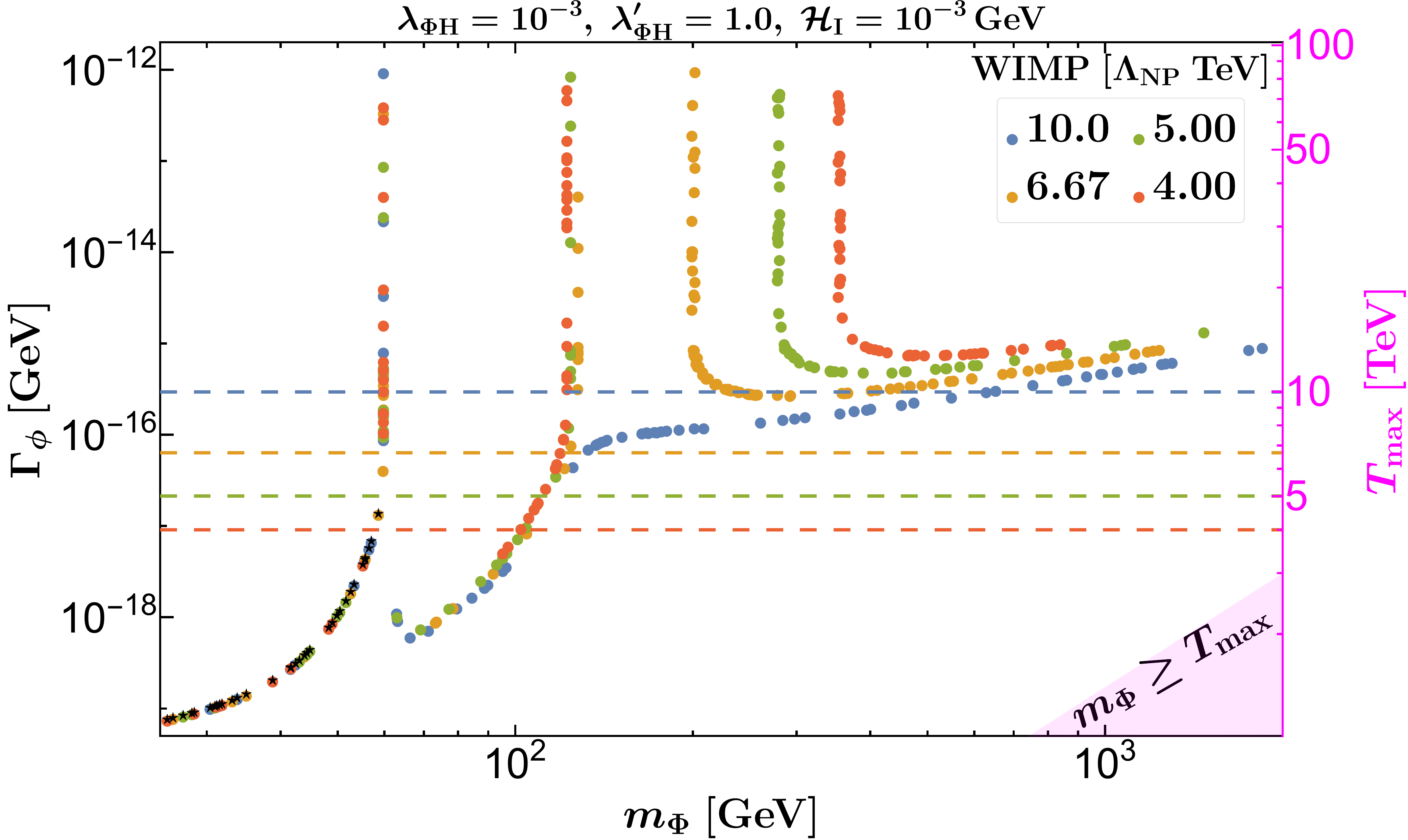}
\caption{
{\textit{Left}: }The different contours depict the parameter space in the $\mphi-\lphiH$ plane along which relic density is satisfied. The different color codes indicate various values of $\lphiHp$. The gray-shaded region is excluded by the $\rm LZ-2025$ data while
the neutrino-floor for direct DM detection \cite{Billard:2021uyg} is represented by the gray dashed line.
{\textit{Right}: }The different contours depicts the relic density allowed parameter space in the $\mphi - \Gphi$ plane, considering the genesis of DM during the reheating era.
The right $y$-axis, depicted with magenta color, indicates the variation of $\Tmax$ with $\Gphi$ (left y-axis), for $\HI = 10^{-3}~\GeV$. The black star points do not satisfy the $\rm LZ-2025$ bound on $\sphin$.
The dashed lines correspond to $\lnp=\Tmax$.
For both plots, we have considered the remaining parameters as $\lphi=1$, and $ \mu_3=2\mphi$.}
\label{fig:relic-comp}
\end{figure}

The relic abundance of DM is evaluated by solving the Boltzmann equation, \eq\eqref{eq:beq}, at the present epoch. We have solved it for two separate scenarios: the thermal freeze-out occurs during (i) the standard radiation-dominated era followed by instantaneous reheating, and (ii) the reheating phase, where the maximum bath temperature is governed by \eq\eqref{eq:tmax}. In the second case, the newly introduced model parameters, $\Gphi$ and $\HI$, determine the values of $\Trh$ and $\Tmax$, which are constrained by CMB and BBN observations, as discussed in the previous section. The parameters $\mphi$, $\lphiH$, $\mut$, $\lphiHp$, $\lnp$, $\Trh$, and $\Tmax$ play a vital role in determining the DM number density, as they dictate the annihilation processes, illustrated in \fig\ref{fig:relic-comp}.
\begin{figure}[htb!]
\centering
\begin{adjustbox}{width=\textwidth}
\begin{tcolorbox}[
colback=gray!5, colframe=black!10, boxrule=0.8pt, arc=4mm, boxsep=0pt, left=0pt, right=0pt, top=0pt, bottom=0pt, width=1\textwidth, halign=center]
\begin{tikzpicture}
\begin{feynman}
\vertex (a);
\vertex[above left=0.7cm and 0.7cm  of a] (a1){\(\Phi\)};
\vertex[below left=0.7cm and 0.7cm  of a] (a2){\(\Phi\)}; 
\vertex[right=0.7cm of a] (b); 
\vertex[above right=0.7cm and 0.7cm of b] (c1){\(\SM\)};
\vertex[below right=0.7cm and 0.7cm of b] (c2){\(\SM\)};
\diagram*{
(a1) -- [line width=0.25mm, charged scalar, arrow size=0.7pt,style=black] (a),
(a) -- [line width=0.25mm, charged scalar, arrow size=0.7pt,style=black] (a2),
(a) -- [line width=0.25mm, scalar,arrow size=0.7pt, edge label'={\(\color{black}{h}\)}, style=black!50] (b),
(b) -- [line width=0.25mm, plain, arrow size=0.7pt] (c1),
(b) -- [line width=0.25mm, plain, arrow size=0.7pt] (c2)};
\node at (a)[circle,fill,style=black,inner sep=1pt]{};
\node at (b)[circle,fill,style=black,inner sep=1pt]{};
\end{feynman}
\end{tikzpicture}
\label{feyn:wimp1}
\begin{tikzpicture}
\begin{feynman}
\vertex (a);
\vertex[above left=0.7cm and 0.7cm of a] (a1){\(\Phi\)};
\vertex[above right=0.7cm and 0.7cm  of a] (a2){\(h\)}; 
\vertex[below left=0.7cm and 0.7cm of a] (c1){\(\Phi\)};
\vertex[below right=0.7cm and 0.7cm of a] (c2){\(h\)};
\diagram*{
(a1) -- [line width=0.25mm, charged scalar, arrow size=0.7pt,style=black] (a),
(a2) -- [line width=0.25mm, scalar, arrow size=0.7pt,style=black] (a),
(a) -- [line width=0.25mm, charged scalar, arrow size=0.7pt] (c1),
(a) -- [line width=0.25mm, scalar, arrow size=0.7pt] (c2)};
\node at (a)[circle,fill,style=black,inner sep=1pt]{};
\end{feynman}
\end{tikzpicture}
\label{feyn:wimp2}
\begin{tikzpicture}
\begin{feynman}
\vertex (a);
\vertex[above left=0.7cm and 0.7cm  of a] (a1){\(\Phi\)};
\vertex[below left=0.7cm and 0.7cm  of a] (a2){\(\Phi\)}; 
\vertex[right=0.7cm of a] (b); 
\vertex[above right=0.7cm and 0.7cm of b] (c1){\(\Phi\)};
\vertex[below right=0.7cm and 0.7cm of b] (c2){\(h\)};
\diagram*{
(a) -- [line width=0.25mm, charged scalar, arrow size=0.7pt,style=black] (a1),
(a) -- [line width=0.25mm, charged scalar, arrow size=0.7pt,style=black] (a2),
(a) -- [line width=0.25mm, charged scalar,arrow size=0.7pt, edge label'={\(\color{black}{\Phi}\)}, style=black!50] (b),
(b) -- [line width=0.25mm, charged scalar, arrow size=0.7pt] (c1),
(b) -- [line width=0.25mm, scalar, arrow size=0.7pt] (c2)};
\node at (a)[circle,fill,style=black,inner sep=1pt]{};
\node at (b)[circle,fill,style=black,inner sep=1pt]{};
\end{feynman}
\end{tikzpicture}
\label{feyn:wimp3}
\begin{tikzpicture}
\begin{feynman}
\vertex (a);
\vertex[above left=0.35cm and 0.7cm of a] (a1){\(\Phi\)};
\vertex[above right=0.35cm and 0.7cm  of a] (a2){\(\Phi \)}; 
\vertex[below = 0.7cm of a] (b); 
\vertex[below left=0.35cm and 0.7cm of b] (c1){\(\Phi\)};
\vertex[below right=0.35cm and 0.7cm of b] (c2){\(h\)};
\diagram*{
(a1) -- [line width=0.25mm, charged scalar, arrow size=0.7pt,style=black] (a),
(a2) -- [line width=0.25mm, charged scalar, arrow size=0.7pt,style=black] (a),
(b) -- [line width=0.25mm, charged scalar, arrow size=0.7pt, edge label={\(\color{black}{\Phi}\)}, style=black!50] (a) ,
(c1) -- [line width=0.25mm, charged scalar, arrow size=0.7pt] (b),
(b) -- [line width=0.25mm, scalar, arrow size=0.7pt] (c2)};
\node at (a)[circle,fill,style=black,inner sep=1pt]{};
\node at (b)[circle,fill,style=black,inner sep=1pt]{};
\end{feynman}
\end{tikzpicture}
\label{feyn:wimp4}
\begin{tikzpicture}
\begin{feynman}
\vertex (a);
\vertex[above left=0.7cm and 0.7cm of a] (a1){\(\Phi\)};
\vertex[above right=0.7cm and 0.7cm  of a] (a2){\(\Phi\)}; 
\vertex[below left=0.7cm and 0.7cm of a] (c1){\(\Phi\)};
\vertex[below right=0.7cm and 0.7cm of a] (c2){\(h\)};
\diagram*{
(a1) -- [line width=0.25mm, charged scalar, arrow size=0.7pt,style=black] (a),
(a2) -- [line width=0.25mm, charged scalar, arrow size=0.7pt,style=black] (a),
(c1) -- [line width=0.25mm, charged scalar, arrow size=0.7pt] (a),
(a) -- [line width=0.25mm, scalar, arrow size=0.7pt] (c2)};
\node at (a)[circle,fill,style=black,inner sep=1pt]{};
\end{feynman}
\end{tikzpicture}
\label{feyn:wimp5}
\end{tcolorbox}
\end{adjustbox}
\caption{The Feynman diagrams correspond to the self- and semi-annihilation of DM, where $\SM=\{\rm h,~W^{\pm},~Z,~quarks~and~leptons\}$.}
\label{fig:feyn-wimp}
\end{figure}
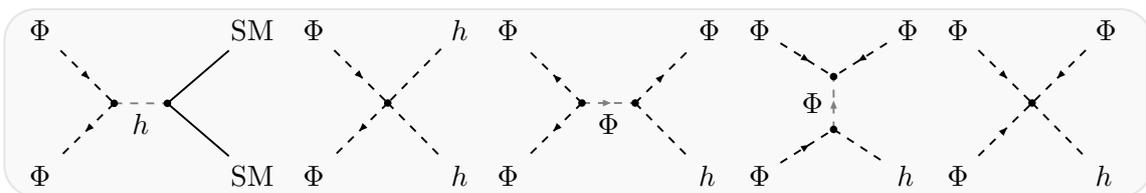
The freeze-out is governed not only by DM self-annihilation but also by semi-annihilation processes, as illustrated in \fig\ref{fig:feyn-wimp}. This is to note that, the new physics (NP) parameters $(\lnp,~\lphiHp)$ of the dim-5 DM–EFT operators contribute exclusively to DM semi-annihilation processes, in addition to the portal $(\lphiH)$ and trilinear $(\mut)$ couplings.
Notably, in this analysis, we consistently respect the EFT validity criterion $2\mphi < \lnp$, assuming CSDM freeze-out near the threshold $\sqrt{s} \simeq 2\mphi$, motivated by the dominance of the s-wave contribution to CSDM semi-annihilation during freeze-out.

The CSDM can interact with the detector nuclei, and the resulting nuclear recoil may serve as a probe for the presence of DM. Numerous experiments \cite{XENON:2025vwd,PandaX:2024qfu,XLZD:2024nsu} are actively searching for DM; however, in this study, we focus solely on the most stringent constraint, the spin-independent DM-nucleon $(\rm \Phi-N)$ scattering cross section $(\sigma_{\rm \Phi N}^{\text{SI}})$, currently reported by the LUX-ZEPLIN experiment \cite{LZ:2024zvo}.
In this model framework, DM interacts with nucleons through the Higgs portal interaction. The parameters relevant for governing the cross-section $\sigma_{\Phi\rm N}^{\text{SI}}$ are, ${\mphi\text{ and }\lphiH}$. This complementarity: the direct detection constraint is governed by $\lambda_{\Phi H}$, while the relic density is governed by additional couplings such as $\lambda^{\prime}_{\Phi H}$ and $\mu_3$ alleviates the severe constraint that a CSDM model with minimal renormalizable interaction faces.

\begin{figure}[htb!]
\centering
\includegraphics[width=0.65\linewidth]{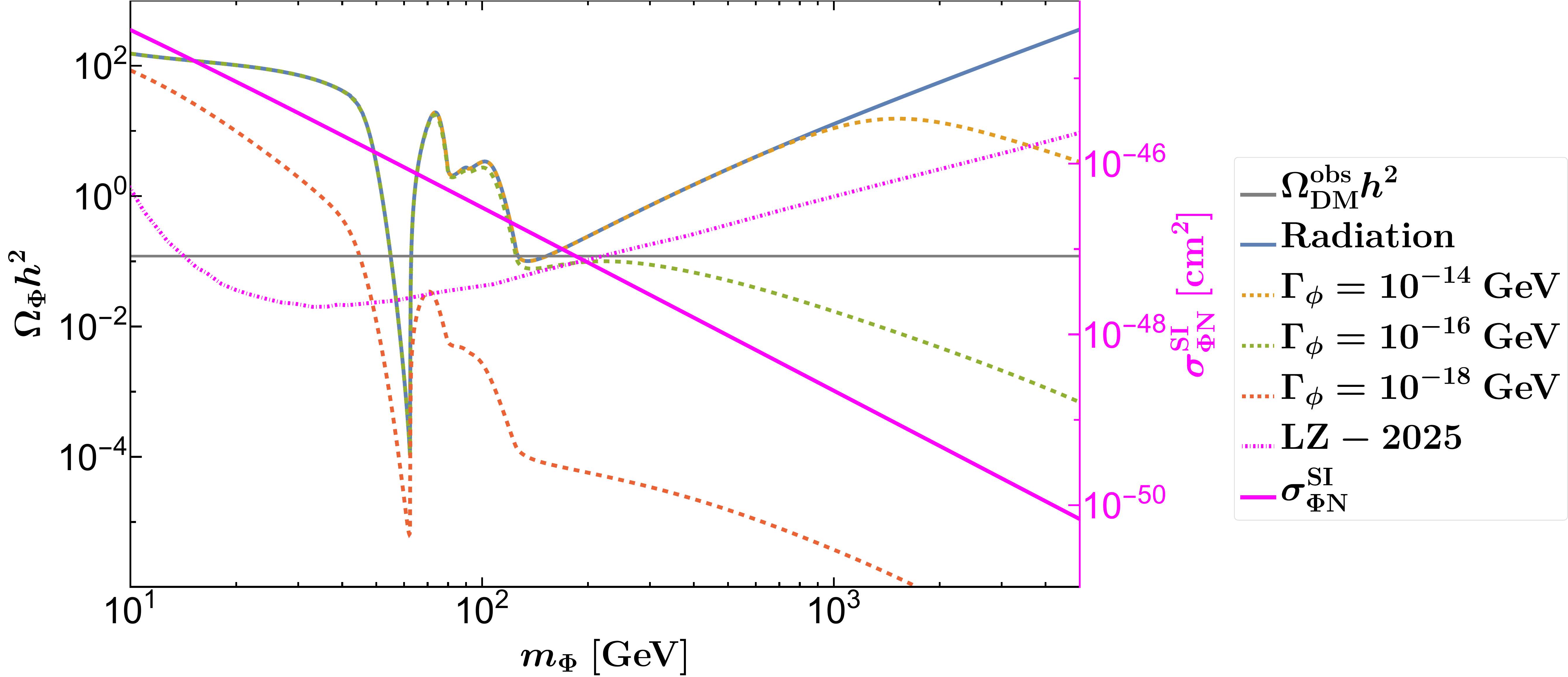}
\caption{
In this figure, we show the variation of the relic density, $\Omega_{\Phi} h^2$, with the DM mass, $\mphi$, for scenarios when freeze-out occurs during the standard radiation domination epoch and during reheating phase. They are, represented by the thick blue and dashed (yellow, green, magenta) lines, respectively.
In addition, the right y-axis in this figure show the DM-nucleon scattering cross section.
The thick magenta line represents the $\Phi$-nucleon scattering cross section.
The dot-dashed magenta line indicating the $\rm LZ-2025$ limit on $\Phi$-nucleon scattering cross section. For this figure, we have considered different parameters as $\lnp=20\,\TeV,~\lphi = 1, ~\mu_3 = 2 \mphi, ~\lphiH =  10^{-2},~\lphiHp = 1$ and $\HI=10^{-4}\,\GeV$.
}
\label{fig:dd1}
\end{figure}

In \fig\ref{fig:dd1}, we show the variation of the DM relic density $(\Omega_{\Phi}h^2)$ and the spin-independent DM-nucleon scattering cross section $(\sphin)$ (magenta y-axis in the right) as functions of the DM mass $\mphi$, while keeping all other parameters fixed as specified in the figure caption. The magenta dot-dashed line represents the upper limit on the spin-independent DM–nucleon scattering cross section from $\rm LZ-2025$.
The results are presented for three values of the inflaton decay width, $\Gphi = \{10^{-14},~10^{-16},~10^{-18}\}$ $\GeV$, represented by the dashed brown, green, and orange lines, respectively. For these three values of $\Gphi$, with $\HI = 10^{-4}~\GeV$, the corresponding reheating and maximum temperatures are $\Trh = \{55.44,~5.70,~0.61\}~\GeV$ and $\Tmax = \{16.23,~5.13,~1.62\}~\TeV$, respectively.
The thick blue line represents the standard radiation-dominated scenario. The green and yellow dashed lines begin to deviate from it when the DM mass is such that the freeze-out temperature $(\TFO)$ falls below the reheating temperature $(\Trh)$ while the deviation for brown line has started at $\mphi\ll 10\,\GeV$.
For the standard freeze-out scenario, when $\mphi > \mh$, the relic density rises progressively with increasing $\mphi$, as both the annihilation and semi-annihilation rates of the CSDM into SM particles diminish with $\mphi$. For $\mphi \lesssim \mh$, we observe four dips in the blue relic-density curve originating from CSDM annihilation into various SM final states, which dominantly contribute to the relic density. From left to right, the four dips arise due to the dominant ($\gtrsim 10\%$) contributions to the relic density from the processes
$\Phi\Phi^* \to b\bar{b}$, $\Phi\Phi^* \to W^+W^-$, $\Phi\Phi^* \to ZZ$, and $\Phi\Phi \to h\,\Phi^*$, respectively. A similar behaviour is observed for the other three dashed relic-density curves; however, it diminishes once the effects of the reheating phase become significant in determining the DM relic abundance.
We note that the DM-nucleon scattering cross section gradually decreases with the enhancement of $\mphi$, and some of the parameter space is excluded by the $\rm LZ-2025$ data, which are mostly overabundant.
Consequently, the $\rm LZ-2025$ allowed parameter space can be used for further studies if the freeze-out is assumed to occur exclusively during the reheating epoch.
\begin{figure}[htb!]
\centering
\subfloat[]{\includegraphics[width=0.49\linewidth]{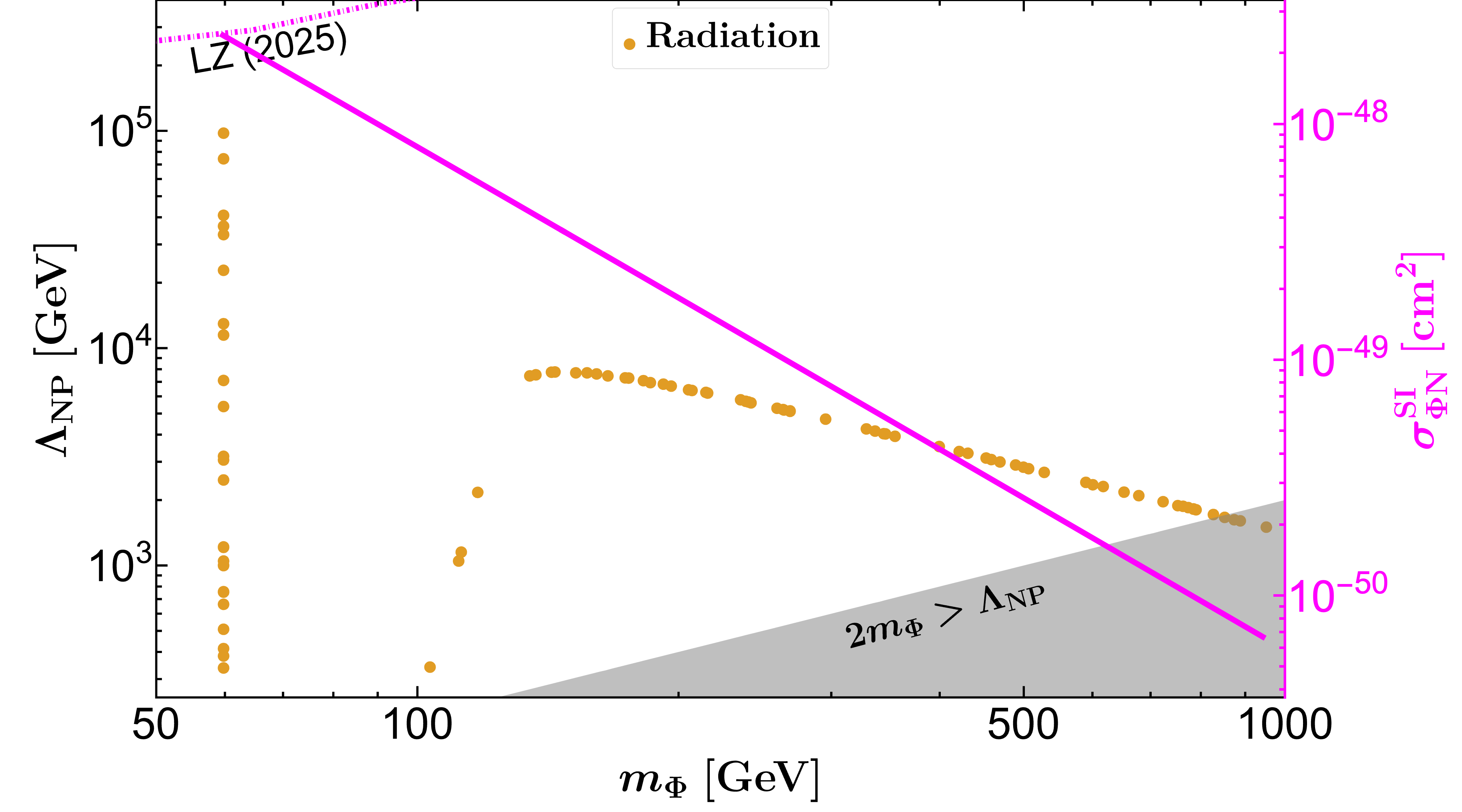}\label{fig:dd2}}~~
\subfloat[]{\includegraphics[width=0.49\linewidth]{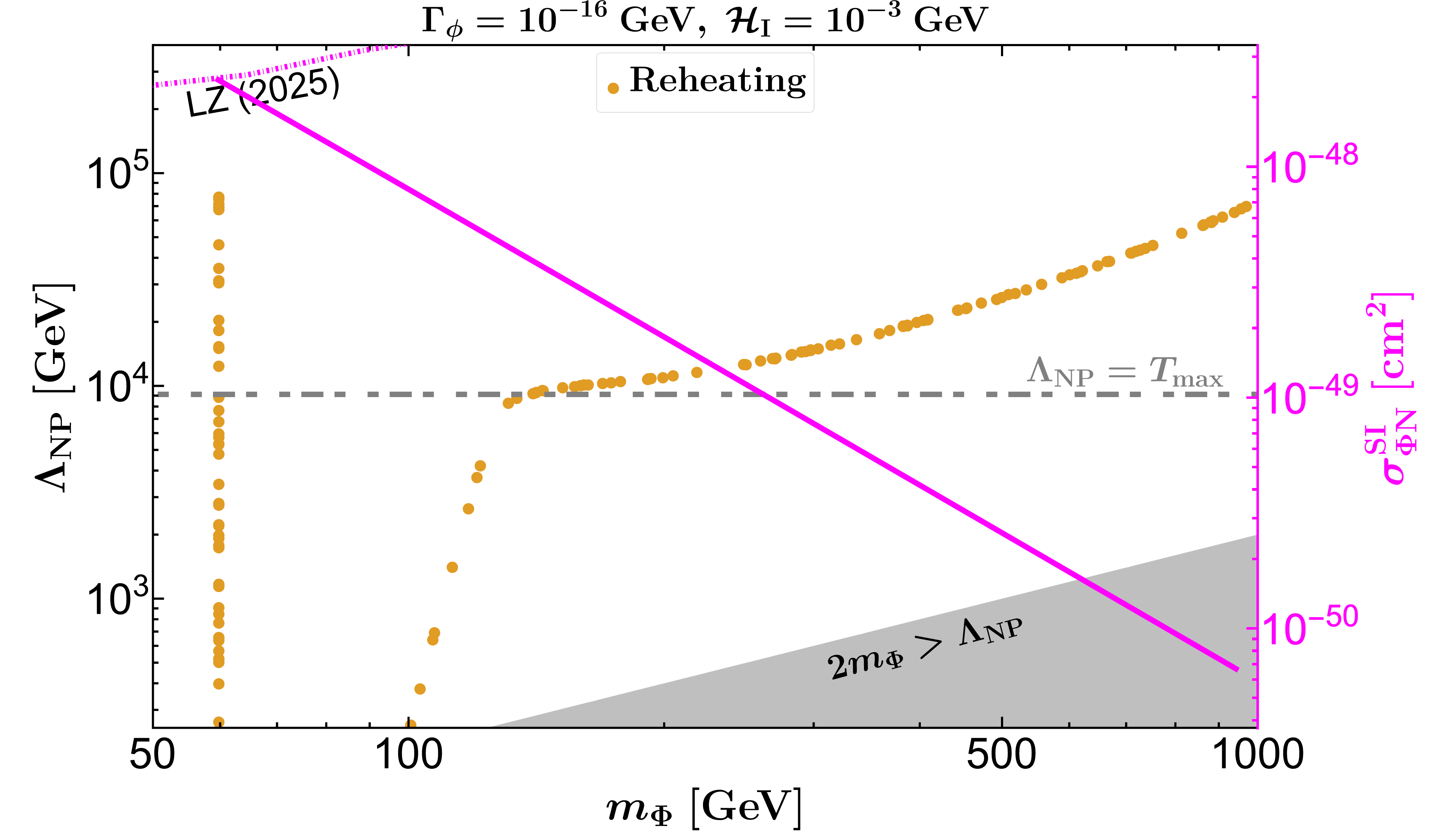}\label{fig:dd3}}
\caption{
The orange points in \fig\ref{fig:dd2} and \ref{fig:dd3} indicate parameter regions consistent with the observed relic density in the $\mphi$–$\lnp$ plane. The magenta color-coded lines correspond to the $\mphi$–$\sphin$ plane (x-axis and right y-axis) in both plots. In the left panel, these points correspond to freeze-out during the radiation-dominated epoch, while in the right panel, they correspond to freeze-out during the reheating epoch.
In both plots, the thick magenta line represents the $\Phi$-nucleon scattering cross section considering parameters as $\lphi = 1, ~\mu_3 = 2 \mphi, ~\lphiH =  10^{-3}$, and $ ~\lphiHp = 1$. In addition, the right y-axis in figures shows the CSDM-nucleon scattering cross section, with the dot-dashed magenta line indicating the $\rm LZ-2025$ limit on DM-nucleon scattering. The gray-shaded region is excluded due to the EFT validity criterion, $\lnp>2\mphi$.}
\label{fig:dd}
\end{figure}

In \fig\ref{fig:dd}, the yellow points indicate parameter values consistent with the observed DM relic density in the $\mphi$–$\lnp$ plane for scenarios in which freeze-out occurs during the radiation-dominated epoch (\fig\ref{fig:dd2}) and the reheating phase (\fig\ref{fig:dd3}). 
Similar to \fig\ref{fig:dd1}, the variation of $\sphin$ is shown on the right y-axis, with the thick magenta line representing its dependence on $\mphi$ and the dot-dashed magenta line indicating the current LZ-2025 direct-detection limit.
As $\lnp$ gradually increases, the semi-annihilation cross section correspondingly decreases, leading to an overproduction of the relic density. This overproduction can be mitigated by selecting a very small $\Trh$ (i.e., $\Gphi$).
With increasing $\mphi$, a decrease in $\lnp$ during the radiation-dominated era (\fig\ref{fig:dd2}) and an increase in $\lnp$ in the reheating scenario (\fig\ref{fig:dd3}) are required to reproduce the observed DM relic density.
It should be noted that the resulting $\Tmax$ must remain below $\lnp$ for the chosen values of $\Gphi$ and $\HI$.
The gray dashed line in \fig\ref{fig:dd3} denotes the lower limit $\lnp = \Tmax$, applicable for the scenario where relic density is predominantly governed by the CSDM semi-annihilation\footnote{The CSDM self-annihilation is independent of the cut-off scale $\lnp$.}. Notably, if the correct relic density is achieved with thermal freeze-out occurring in the radiation-dominated era ($\Trh \gg \TFO$) where the processes govering the relic density is $\lnp$ independent, $\lnp > \Tmax$ may be relaxed.

In the left panel of \fig\ref{fig:relic-comp}, we show the points consistent with the observed DM relic density in the $\mphi$–$\lphiH$ plane, where different colors correspond to different values of $\lphiHp$. The NP cut-off scale has been set to $\Lambda_{\rm{NP}}=2$ TeV, as indicated at the top of the figure.
The effect of the $\lphiHp$ coupling is most significant near the Higgs mass, as it contributes solely to the DM semi-annihilation cross section, $\langle\sigma v\rangle_{\Phi\Phi\to\Phi^* h} \propto \lphiHp$. Its impact becomes particularly pronounced when $\lphiH$ is comparatively small. Therefore, with the enhancement of $\lphiHp$, the DM semi-annihilation cross section increases, and consequently, the relic density decreases. However, this effect becomes diluted when $\mphi \gtrsim 2~\TeV$, and the behavior once again starts to resemble that of the renormalizable CSDM framework.

In the right panel of \fig\ref{fig:relic-comp}, we present the DM relic density allowed points in the $\mphi-\Gphi$ plane, while noting that entropy dilution can modify the relic density. We also show the variation of $\Tmax$, along with $\Gphi$ while $\HI=10^{-3}\,\GeV$, on the right $y$-axis. In this analysis, all model parameters are fixed as specified in the figure inset, except for the NP scale, $\lnp$.
The Burnt Orange, Mustard Yellow, Olive Green, and Steel Blue points correspond to $\lnp = \{4,~5,~6.67,~\text{and}~10.0\}~\text{TeV}$, respectively, and the correct relic density is achieved via the freeze-out mechanism.
However, within a single-color funnel around the Higgs mass, the relic density is significantly under-abundant due to semi-annihilation. Beyond the Higgs mass (i.e., $\mphi \gg \mh$), this effect becomes more pronounced. As $\lnp$ gradually increases, the cross section for semi-annihilation decreases; in this scenario, the relic density can be tuned by reducing the inflaton decay width $\Gphi$, which effectively lowers $\Trh$.
As mentioned in figure caption, the dashed lines correspond to $\lnp=\Tmax$, above which the EFT validity is violated. This constraint is relaxed in the resonant regions $\mphi\sim \mh/2$ as self-annihilation processes are independent of $\lnp$. Notably, for $\lnp=10\,\TeV$, the effect of reheating for obtaining the correct relic density is most pronounced.
The magenta-shaded region in the bottom-right of the right panel denotes the exclusion of thermally produced CSDM by the requirement $\Tmax \geq \mphi$; otherwise, CSDM production occurs non-thermally.
Additionally, the black star points disrespect the current direct detection limits set by $\rm LZ-2025$.
\subsection{Indirect detection}
An alternative approach to probe DM involves the detection of SM particles such as gamma rays, neutrinos, and cosmic rays (electrons, positrons, antiprotons), that are generated due to DM annihilation or decay. Excesses of these particles in regions of high DM density, such as the Galactic Center, dwarf spheroidal galaxies, and galaxy clusters, could provide indirect evidence for DM. Experiments targeting these signals include Fermi-LAT \cite{Fermi-LAT:2015att, Fermi-LAT:2016afa, Fermi-LAT:2017opo, Fermi-LAT:2025fst, Fermi-LAT:2025gei}, MAGIC \cite{MAGIC:2016xys, MAGIC:2024tnf}, H.E.S.S. \cite{HESS:2020zwn}, and CTA \cite{CTA:2020qlo, Silverwood:2014yza} for gamma rays; IceCube \cite{IceCube:2023ies} and ANTARES \cite{ANTARES:2019svn, ANTARES:2020leh, ANTARES:2015vis} for neutrinos; and AMS-02 \cite{Kounine:2023glo, AMS:2019rhg, Caroff:2016abl} and CALET \cite{CALET:2023emo} for charged cosmic rays. Although no confirmed detections have been reported, these searches impose stringent upper limits on DM annihilation and decay rates into SM particles. Also, the gamma-ray observations in the direction of Dwarf Spheroidal Galaxies (Fermi-LAT) and the Galactic Center (H.E.S.S.), together with Planck measurements of the CMB, place upper limits on DM semi-annihilation \cite{Queiroz:2019acr}, with additional sensitivity expected from CTA.

\begin{figure}[htb!]
\centering
\includegraphics[width=1\linewidth]{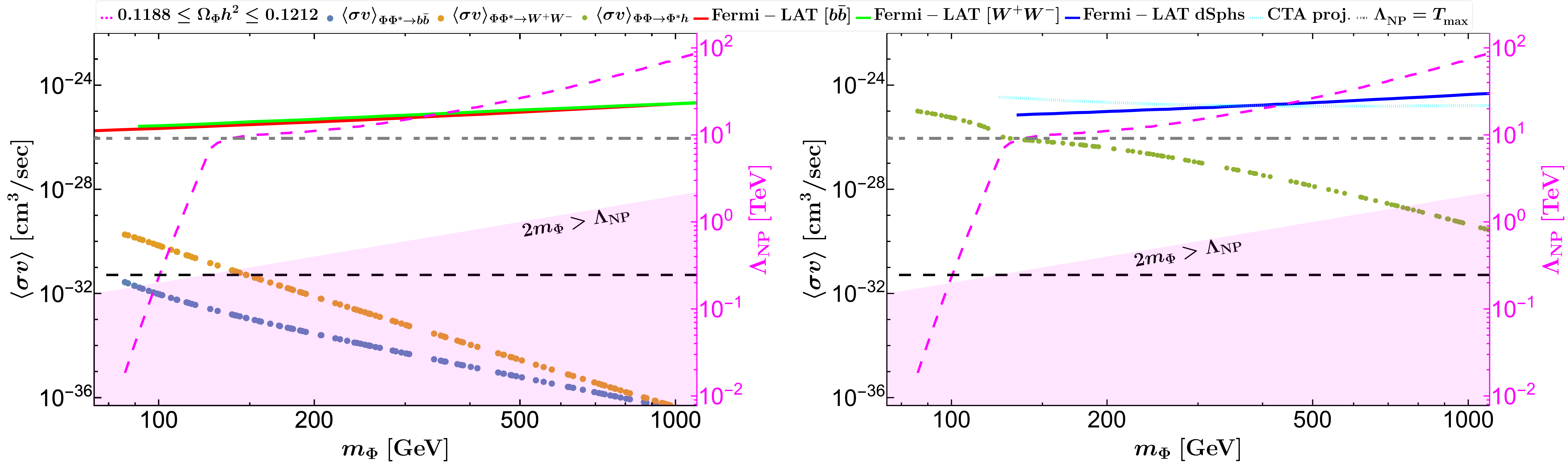}
\caption{In the left and right panels of this figure, we show the variation of DM self- and semi-annihilation into SM particles (depicted via dotted lines) as a function of $\mphi$, respectively. The different parameters are fixed at $\lphi=1$, $\mu_3=2\mphi$, $\lphiH=10^{-3}$, $\lphiHp=1$, $\Gphi=10^{-16}~\GeV$, and $\HI=10^{-3}~\GeV$.
The magenta color–coded right y-axes represents the $\lnp$ scale.
The blue and yellow points in the left panel correspond to the self-annihilation of CSDM into $b\bar{b}$ and $W^+W^-$, respectively, while the green points in the right panel corresponds to the semi-annihilation of CSDM into $\Phi^* h$.
The cyan dotted (CTA) and blue thick (Fermi-LAT) lines, along with the red and green thick Fermi-LAT lines, represent the projected and observed upper limits on the DM self- and semi-annihilation cross sections, respectively, as shown in the figure legend.
The magenta long dashed lines show the relic-allowed parameter space in the $\mphi-\lnp$ plane, with the variation of $\lnp$ indicated along the right y-axis.
All magenta points below the gray dot-dashed line are excluded by the EFT validity condition $\lnp > \Tmax$, relevant for semi-annihilation, while the relic-density–allowed magenta points within the magenta-shaded region are disfavoured by $2\mphi < \lnp$.
The olive points must satisfy $2\mphi < \lnp$ due to the $\lnp$ dependence of the semi-annihilation cross section.
The black dashed line represents $\lnp=v_{\rm SM}^{}$ \cite{Buchmuller:1985jz,Brivio:2017vri,Grzadkowski:2010es}, where $v_{\rm SM}^{}$ denotes the EWSB vev.}
\label{fig:relic_ID}
\end{figure}
In this work, we show the constraint on the self-annihilation of the CSDM via the Higgs-portal interaction into SM final states ($b\bar b$ and $W^+W^-$) using the stringent limits from the Fermi-LAT experimental data. In addition, we also show the constraint on the semi-annihilation channel $\Phi\Phi \to \Phi^{*} h$. To derive these constraints, we use the limits obtained in ref\,. \cite{Queiroz:2019acr} and perform a rescaling, as we have a CSDM.
The self-annihilation cross section into $b\bar b$ and $W^+W^-$ decreases with increasing $\mphi$ and decreasing $\lphiH$. Consequently, choosing $\lphiH = 10^{-3}$ helps render the parameter space consistent with the Fermi-LAT bounds, as also seen in \fig\ref{fig:relic_ID}, where the blue and yellow points correspond to DM annihilation into $b\bar b$ and $W^+W^-$, respectively. In contrast, the semi-annihilation cross section depends not only on $\lphiH$ but also on $\mu_3$, $\lphiHp$, and $\lnp$. While larger values of $\mu_3$ and $\lnp$ aid in achieving the correct relic density for a given $\lphiH$, they are more tightly constrained by the Fermi-LAT limits on the semi-annihilation cross section. As shown in \fig\ref{fig:relic_ID}, some part of the parameter space is excluded by the Fermi-LAT bounds (thick blue) and the CTA projections (dashed cyan), even though they remain consistent with direct-detection and relic-density constraints.
Foremost, the condition $\lnp > \Tmax$ is relevant only for the relic-density analysis during the reheating era, whereas $\lnp > 2\mphi$ is required to estimate the semi-annihilation rate dominated by the NP coupling. In contrast, self-annihilation into SM particles is unaffected by either condition during the radiation-dominated epoch. Accordingly, in the right (left) panel of \fig\ref{fig:relic_ID}, we show only the semi-annihilation into $\Phi^{*} h$ (self-annihilation into $b\bar b$ and $W^+W^-$ \cite{Fermi-LAT:2025gei}). Some magenta points below the dashed gray exclusion line remain viable because, for smaller $\lnp$ coupling values, semi-annihilation dominates and yields the observed relic abundance during the radiation-dominated epoch.

\begin{figure}[htb!]
\centering
\includegraphics[width=0.49\linewidth]{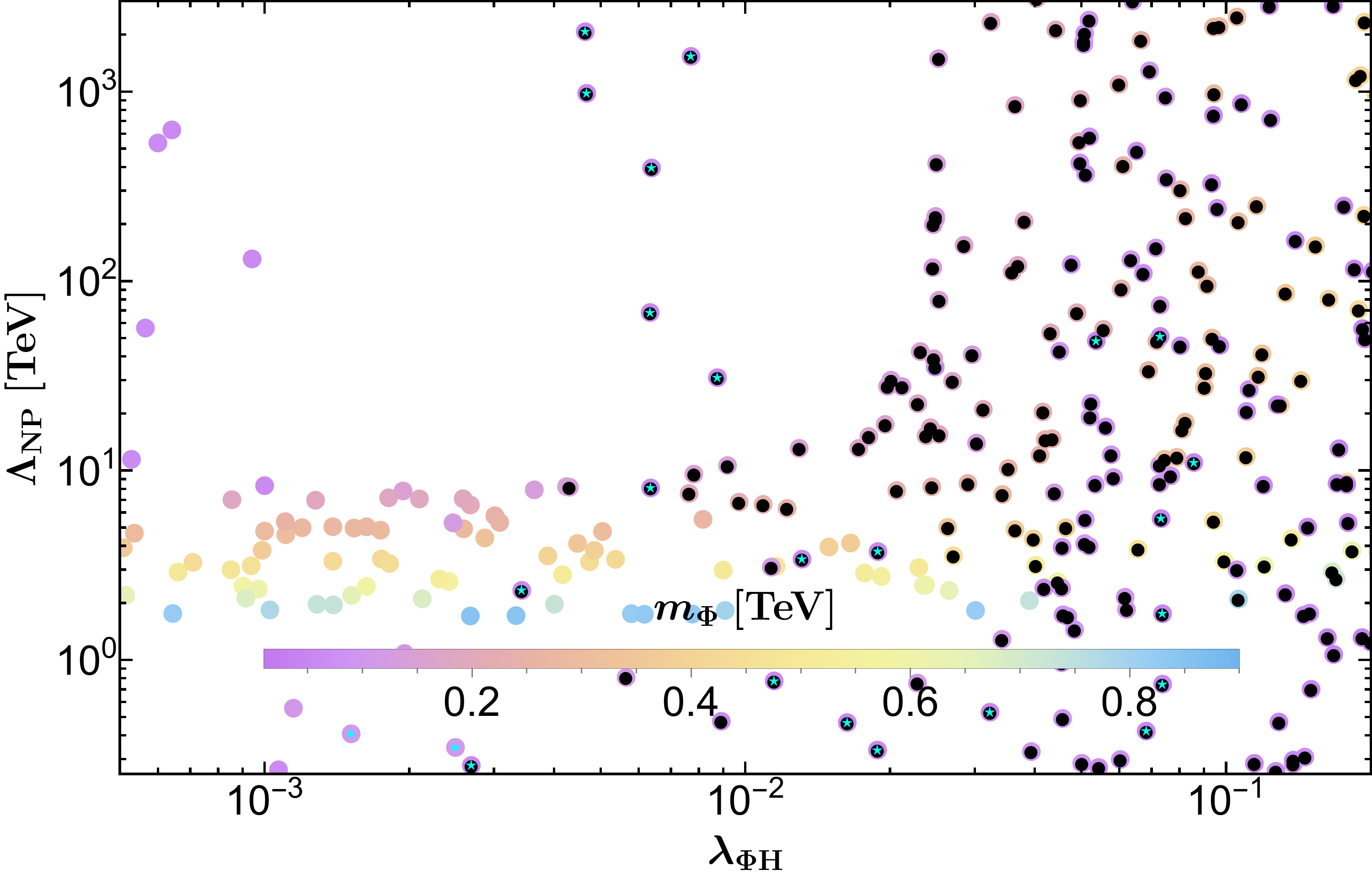}~~
\includegraphics[width=0.49\linewidth]{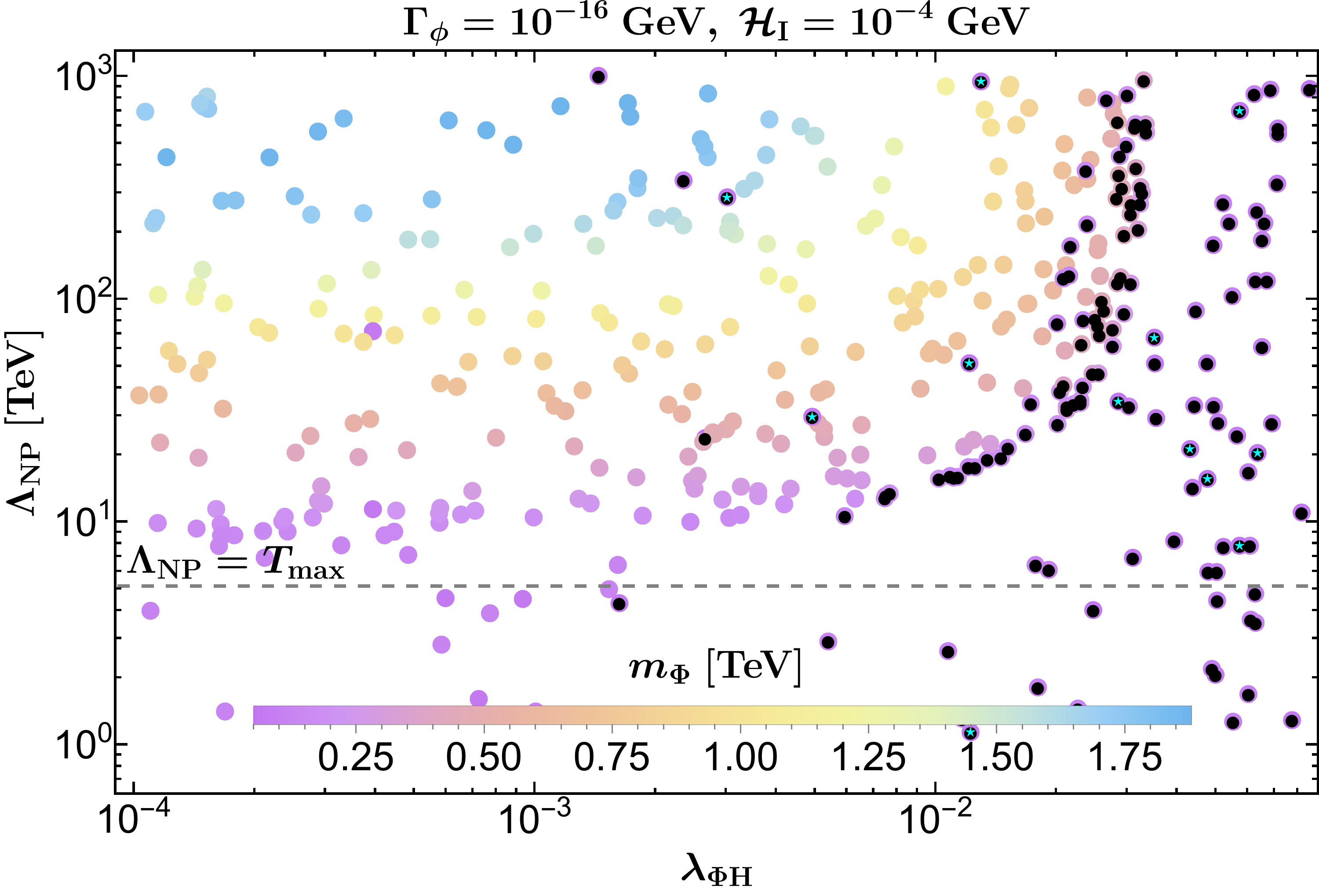}
\caption{The relic density allowed parameter space is shown in the $\lphiH$–$\lnp$ plane, with the pastel color bar indicating the variation of $\mphi$. The remaining parameters in both plots are fixed at $\lphi=1.0,~\mu_3=\mphi$, and $\lphiHp=1.0$ throughout the scan. The left and right plots correspond to the genesis of DM during radiation domination and reheating era (with ${\Trh=5.70~\GeV,\text{ and }\Tmax=5.13~\TeV}$), respectively. The black points overlaid on the pastel-colored points are excluded by the stringent exclusion limit on $\sigma_{\Phi\rm N}^{\rm SI}$ using $\rm LZ-2025$ data. All cyan five-star points overlaid on the black points are excluded by the Fermi-LAT upper limits on dark-matter self-annihilation, $\Phi \Phi^* \to b\bar{b}~(W^+W^-)$, and semi-annihilation, $\Phi \Phi \to \Phi^* h$, cross sections. The gray dashed line corresponds to $\lnp = \Tmax$.
}
\label{fig:scan}
\end{figure}

Finally, after incorporating all relevant bounds, we present in \fig\ref{fig:scan} a summary of the parameter space scan in the $\lphiH-\lnp$ plane. The pastel color bar illustrates the variation of $\mphi$ for two distinct scenarios.
In the left panel, freeze-out is assumed to occur during the radiation-dominated era ($\Trh \gg \TFO$), while in the right panel it might take place during the reheating epoch ($\TFO\gg\Trh$), with $\Trh = 5.70~\GeV$ and $\Tmax = 5.13~\TeV$ for the chosen values of $\Gphi = 10^{-16}~\GeV$ and $\HI = 10^{-4}~\GeV$.

In both plots of \fig\ref{fig:scan}, the pastel-colored points satisfy relic-density constraints, and also respect the criteria $\lnp>2\mphi$. Additionally, in the right plot we also show the line $\lnp=\Tmax$. The black points superimposed on the pastel regions violate the $\rm LZ-2025$ direct-detection bound, while the cyan five-star points overlaid on either pastel or black points fail to meet the Fermi-LAT limits on DM self- and semi-annihilation cross sections.
In the left panel, after imposing all relevant bounds on CSDM, allowed points remain near the Higgs resonance, around the Higgs mass, and for $\mphi \gtrsim 1~\TeV$, although this comes at the cost of the Higgs-portal coupling $\lphiH$.
The top-left white region corresponds to points that do not reproduce the correct relic abundance and instead lead to an overabundant DM relic density. This region can, however, be reconciled with the observed abundance if the freeze-out occurs during reheating, as we show in the right panel of \fig\ref{fig:scan}.
As is evident from the figure, a wide range of DM mass is allowed from the relic density as well as direct and indirect experimental constraints.
For our chosen parameters, listed in the \fig\ref{fig:scan} caption, we have $\Tmax = 5.13~\TeV$, which allows a new-physics scale $\lnp > 5.13~\TeV$.
Although we adopt $\Trh = 5.70~\GeV$ for $\Gphi = 10^{-16}~\GeV$ and $\HI = 10^{-4}~\GeV$ in this setup, the allowed parameter space may shift for different choices of these inputs.

In conclusion, once relic-density, direct, and indirect bounds are imposed, the allowed points correspond to $\lphiH \lesssim 10^{-2}$, spanning a wide range of CSDM masses, with a new-physics scale $\lnp \gtrsim 1~\TeV$, while lower values may be accessible under appropriate reheating dynamics.


\subsection{Collider search}
In this section, we show the production cross sections of the DM signal at future collider experiments \cite{ZurbanoFernandez:2020cco, FCC:2025uan}. We begin by summarizing in \cref{tab:benchmark} a set of benchmark points (BPs) that simultaneously satisfy the relic-density requirement and are consistent with current constraints from direct and indirect DM searches. Benchmark points \textbf{BP1}, \textbf{BP2}, and \textbf{BP3} correspond to scenarios with radiation domination, while \textbf{BP4} is for the reheating-dominated era. In the subsequent analysis, we focus on parameter choices similar to \textbf{BP1}. 

\begin{table}[htb!]
\centering
\renewcommand{\arraystretch}{1.0} 
\resizebox{\textwidth}{!}{%
\begin{tabular}{|c|c|c|c|c|c|c|c|c|}
\hline
\rowcolor{gray!25} Benchmark points & $\mphi[\GeV]$ & $\mu_3[\GeV]$ & $\lphiH$ & $\lnp[\GeV]$&$\sphin[\rm cm^2]$ &$\Trh[\GeV]$ \\
\hline
\rowcolor{cyan!20} BP1 & $112$ & $112$ & $ 0.002$ & $1081$& $2.64\times 10^{-48}$&$-$ \\
\rowcolor{cyan!20} BP2 & $59.7$ & $59.7$ & $ 0.001$  & $263$& $2.74\times 10^{-48}$ &$-$ \\
\rowcolor{cyan!20} BP3 & $106.9$ & $106.9$ & $ 0.001$ & $556$& $9.99\times 10^{-49}$ & $-$\\
\rowcolor{cyan!20} BP4 & $112.72$ & $112.72$ & $ 0.001$ & $1403$& $3.46\times 10^{-49}$ &$5.70$\\
\hline
\end{tabular}}
\caption{All benchmark points (BPs) satisfy the relic-density constraint and the bounds from direct, indirect, and collider searches for DM. We take $\lphiHp = 1.0$, and for BP4 the corresponding value of $\Tmax$ is $5.13\,\TeV$.}
\label{tab:benchmark}
\end{table}

\begin{figure}[htb!]
\centering
\includegraphics[width=0.60\linewidth]{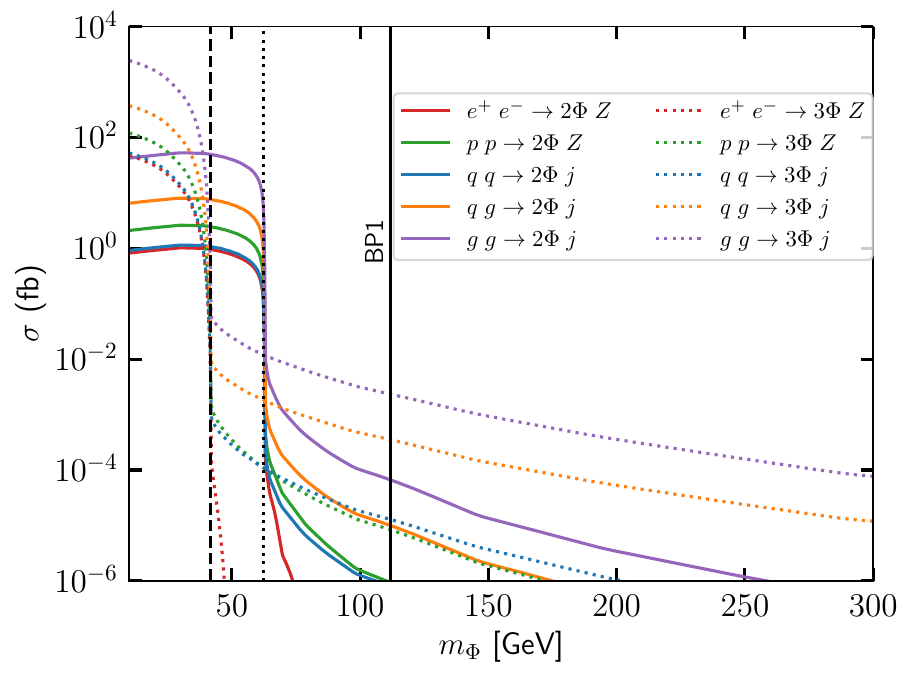}
\caption{The CSDM production cross section ($\sigma$) as a function of $\mphi$ at the FCC-ee with $\sqrt{s} = 240~\GeV$ and at HL-LHC with $\sqrt{s} = 14~\TeV$, shown for both $2\Phi$ and $3 \Phi$.
In the analysis, we adopt the \textbf{BP1} parameter values to evade current bounds from direct detection ($\rm LZ-2025$ limit on the spin-independent DM–nucleon cross section) and indirect detection (Fermi-LAT limits on DM annihilation into $W^+W^-$ and $b\bar b$).
The vertical thick black line indicates $\mphi=112\,\GeV$, as listed in \cref{tab:benchmark}, while the vertical dashed and dotted black lines pinpoint $\mphi=\mh/2$ and $\mphi=\mh/3$, respectively.}
\label{fig:fccee-hllhc-xsec}
\end{figure}

In \fig\ref{fig:fccee-hllhc-xsec}, we show the production cross sections for various $\Phi$ production channels at the HL-LHC and FCC-ee, corresponding to centre-of-mass energies of $\sqrt{s} = 14~\text{TeV}$ and $\sqrt{s} = 240~\text{GeV}$, respectively, as functions of the dark scalar mass $m_{\Phi}$. The solid and dotted curves correspond to the production of $2\Phi$ and $3\Phi$, respectively, in association with different particles at the hadron and lepton colliders, as indicated by different colours. The red and green curves represent the associated production of $\Phi$ with a $Z$ boson at FCC-ee and HL-LHC, respectively. The blue, orange, and violet curves correspond to associated production with one jet at the HL-LHC, originating from different partonic initial states, namely quark--quark $(qq)$, quark--gluon $(qg)$, and gluon--gluon $(gg)$ initial states, respectively. The black dashed and dotted vertical lines indicate the CSDM mass values $m_{\Phi} = m_{h}/3$ and $m_{\Phi} = m_{h}/2$, respectively, while the thick black line marks the CSDM mass of BP1 given in \cref{tab:benchmark}.

For the red solid and dotted curves, corresponding to FCC-ee processes, the cross sections drop to nearly zero at approximately $m_{\Phi} \simeq (\sqrt{s}-m_{Z})/2 \approx 75~\text{GeV}$ and $m_{\Phi} \simeq (\sqrt{s}-m_{Z})/3 \approx 50~\text{GeV}$, respectively. This sharp suppression is purely kinematic in origin and arises due to the lack of available phase space for the on-shell production of $\Phi$ beyond these thresholds. 
In contrast, for the $pp$ collider case, represented by the green, blue, orange, and violet curves, a sudden drop in the cross section is observed at $\mphi = \mh/2$ (indicated by the black dotted vertical line). This feature is a direct consequence of the fact that the on-shell decay $h \rightarrow 2\Phi$ becomes kinematically forbidden beyond this mass threshold. A similar drop in the dotted curves is seen at $\mphi = \mh/3$ (black dashed vertical line), which can be attributed to the same kinematic constraint affecting the corresponding $3\Phi$ production modes.

As observed, for both the $2\Phi\,j$ and $3\Phi\,j$ final states, the gluon--gluon, quark--gluon, and quark--quark initiated subprocesses follow a clear hierarchy in their production cross sections. This ordering directly reflects the corresponding hierarchy of the parton distribution functions, with the gluon PDF dominating over the quark PDFs in the relevant Bjorken-$x$ range \cite{Bailey:2020ooq, McGowan:2022nag, PDF4LHCWorkingGroup:2022cjn}. Consequently, the inclusive contributions to the processes $pp \to 2/3\Phi\, j$ are significantly larger than those of $pp \to 2/3\Phi\, Z$, primarily due to the reduced phase-space suppression in the $Z$-associated production channels.

As can be seen, for benchmark point \textbf{BP1}, the effective number of signal events expected at the HL-LHC is well below one for all considered channels, except for the process $gg \to 3\Phi \;j$. Even in this case, the expected signal yield is at most of order $\mathcal{O}(10)$ events, rendering a signal discovery with a conventional cut-based analysis highly challenging. More sophisticated approaches based on machine-learning (ML) techniques may significantly enhance the sensitivity of such searches; however, a detailed investigation of these methods lies beyond the scope of the present work and is deferred to a future study.

\section{Summary and Conclusion}
\label{sec:summary}
The minimal renormalizable Higgs-portal real (complex) scalar DM model, stabilized by a $\mathcal{Z}_2~(\mathcal{Z}_3)$ symmetry, is now strongly constrained by direct, indirect, and collider searches. Among these probes, direct-detection experiments impose the most stringent bounds on the Higgs-portal coupling. This tension can be mitigated by extending the minimal renormalizable setup with additional dark-sector states. Alternatively, an EFT-based extension involving DM and SM fields and/or a modification of the standard cosmological history also serve as a viable framework to evade this constraint. 

In this work, we examined the phenomenology of a complex scalar DM candidate $\Phi$ stabilized by a $\mathcal{Z}_3$ symmetry, where we extend the renormalizable CSDM model with a dim-5 operator $\mathcal{O}_5$ involving DM field and SM Higgs field (while noting that the analogous extension for real scalar DM first appears at dim-6). In addition to this, the model also contains a self-interaction term involving $\Phi$ field, which however do not play any role in our analysis. Throughout this work, we assumed that the other higher-dimensional operators remain subdominant relative to this leading contribution.
After electroweak symmetry breaking, the $\mathcal{O}_5$ operator generates an additional Higgs-portal interaction $\Phi^3 h$. While this interaction is unconstrained by DM direct-detection search, however this plays a dominant role in setting the relic abundance through semi-annihilation. The free parameters associated with $\mathcal{O}_5$, such as, the new-physics scale and its coupling, are further constrained by Fermi-LAT data on the semi-annihilation cross section.
Additionally, we investigated the freeze-out of the complex scalar DM during the reheating epoch, where regions that are associated with DM overabundance in the radiation-dominated era may instead provide the correct relic density. Notably, the requirement of EFT validity imposes stringent condition on DM mass, the new physics scale of the theory and the maximum temperature of the thermal bath.

After imposing the aforementioned constraints, the remaining viable parameter space lies predominantly near the Higgs resonance, with the CSDM mass close to the Higgs mass and extending above $(1~\TeV)$ in the standard radiation dominated scenario. This restriction can, however, be alleviated if the CSDM freeze-out occurs during reheating epoch. The corresponding allowed region may be probed by future direct and indirect detection experiments. In contrast, searches for CSDM at hadron or lepton colliders remain challenging due to less number of events. Consequently, a substantially refined search strategy will be required to achieve observable signal significance, which will be explored in forthcoming work.

\section*{Acknowledgement}
DP extends his gratitude to Dr. Basabendu Barman for his valuable discussions and insightful comments.



\bibliographystyle{JHEP} 
\bibliography{biblio.bib}
\end{document}